\documentclass[12pt,eqsecnum,draft]{article}
\usepackage[dvips]{graphicx}
\usepackage{amssymb}
\usepackage{amsmath}
\usepackage{ulem}
\usepackage[dvipsnames]{xcolor}
\usepackage{soul}
\usepackage{ascmac}
\usepackage{cite}
\usepackage{url}

\makeatletter

\@addtoreset{equation}{section}
\makeatletter

\newcommand{\qed}{\hbox{\rule[-2pt]{6pt}{6pt}}}
\newcommand{\D}{{\rm d}}

\newtheorem{Prop}{Proposition}

\newtheorem{lm}{Lemma}

\newtheorem{Coro}{Corollary}

\newcommand{\dalm}{\kern1pt\vbox{\hrule height 0.9pt\hbox{\vrule width
0.9pt\hskip 2.5pt\vbox{\vskip 5.5pt}\hskip 3pt\vrule width 0.3pt}\hrule height
0.3pt}\kern1pt}

\begin{document}

\begin{titlepage}
\vfill
\begin{flushright}
\today
\end{flushright}

\vfill
\begin{center}
\baselineskip=16pt
{\Large\bf
Criteria for energy conditions
}
\vskip 0.5cm
{\large {\sl }}
\vskip 10.mm
{\bf Hideki Maeda${}^{a}$ and Tomohiro Harada${}^{b}$} \\

\vskip 1cm
{
${}^a$ Department of Electronics and Information Engineering, Hokkai-Gakuen University, Sapporo 062-8605, Japan.\\
${}^b$ Department of Physics, Rikkyo University, Toshima,
Tokyo 171-8501, Japan. \\
\texttt{h-maeda@hgu.jp, harada@rikkyo.ac.jp}

}
\vspace{6pt}
\end{center}
\vskip 0.2in
\par
\begin{center}
{\bf Abstract}
\end{center}
\begin{quote}
In model building studies, it is important to check the energy conditions for the corresponding energy-momentum tensor determined by the gravitational field equations in order to single out physically reasonable models.
In this process, one often encounters a situation where the energy-momentum tensor has one off-diagonal ``space-time'' component in the frame with an orthonormal basis in a given spacetime.
We derive useful criteria of energy-momentum tensors for their Hawking-Ellis types and the standard energy conditions in such situations.
As demonstrations, we apply those criteria to four different systems.
\vfill
\vskip 2.mm
\end{quote}
\end{titlepage}




\tableofcontents

\newpage

\section{Introduction}

The energy conditions on the energy-momentum tensor of a matter field guarantee that the matter field behaves in a physically reasonable manner.
For example, the weak energy condition (WEC) guarantees that the energy density measured by an observer moving along any timelike orbit is non-negative, and the dominant energy condition (DEC) guarantees that, in addition to the WEC, the energy flux of the matter field does not propagate faster than the speed of light.
Actually, fundamental results in general relativity showing the desirable behaviour of spacetime, such as the black-hole area theorem~\cite{Hawking:1971vc} and the positive mass theorem~\cite{Schon:1979rg,Schon:1981vd,Nester:1982tr,Witten:1981mf}, have been proved under certain energy conditions.
Also, Penrose's singularity theorem~\cite{Penrose:1964wq} has been proved under the null convergence condition, which is equivalent in general relativity to the null energy condition (NEC), the weakest among all the standard energy conditions.

Although a matter field including quantum effects such as the Casimir effect~\cite{Klimchitskaya:2006rw,Klimchitskaya:2009cw} can violate the NEC, it is reasonable to assume that classical matter fields satisfy the energy conditions.
For example, fundamental fields such as an electromagnetic field and a scalar field with an appropriate potential satisfy all standard energy conditions.
A perfect fluid, which is a phenomenological matter field, obeying an appropriate equation of state also satisfies all standard energy conditions.
These results are completely general and have been derived without any assumption of spacetime symmetry~\cite{Maeda:2018hqu}.

However, we sometimes need to study the corresponding matter fields for a given spacetime configuration.
This is the case when we construct a model of a non-singular black-hole spacetime~\cite{Maeda:2021jdc} or a black hole in the expanding universe~\cite{Thakurta1983,Sultana:2005tp,McClure:2006kg,Culetu:2012ih,Mello:2016irl,Harada:2021xze}.
Then, the components of the corresponding energy-momentum tensor $T_{\mu\nu}$ are determined by the gravitational field equations.
In such a case, the energy conditions may be used to check whether the resulting $T_{\mu\nu}$ is physically valid or not.

Energy-momentum tensors $T_{\mu\nu}$ can generally be classified into four Hawking-Ellis types in arbitrary $n(\ge 3)$ dimensions~\cite{he1973,Santos:1994cs,srt1995,hrst1996,bh2002,rst2004}.
In two dimensions ($n=2$), it is classified into three types.
The components of $T_{\mu\nu}$ in an orthonormal frame can be written in a canonical form for each type by a local Lorentz transformation, and inequalities equivalent to the standard energy conditions are available for those orthonormal-frame components.
However, with a natural set of basis vectors or one-forms in a given spacetime, those inequalities cannot be used immediately, as the components in the frame do not take canonical forms in general.
Therefore, the inequalities equivalent to the standard energy conditions for more general orthonormal-frame components would be useful for future investigation.

In the present paper, we derive criteria for the Hawking-Ellis types and the standard energy conditions for the case where the energy-momentum tensor $T_{\mu\nu}$ has a single off-diagonal ``space-time'' component in an orthonormal frame, which is frequently encountered in model building studies.
The organization of the present paper is as follows.
In Sec.~\ref{sec:main}, we will present our main results.
In Sec.~\ref{sec:applications}, we will apply our results to four different systems.
We will summarize our results in the final section.
Our conventions for curvature tensors are $[\nabla _\rho ,\nabla_\sigma]V^\mu ={R^\mu }_{\nu\rho\sigma}V^\nu$ and $R_{\mu \nu }={R^\rho }_{\mu \rho \nu }$.
We adopt units such that $c=1$.
The signature of the Minkowski spacetime is $(-,+,\ldots,+)$, and Greek indices run over all spacetime indices.
Other types of indices will be specified in the main text.

\section{Criteria for energy conditions}
\label{sec:main}
\subsection{Preliminaries}
In an $n(\ge 2)$-dimensional spacetime, the components of an energy-momentum tensor $T_{\mu\nu}$ in an orthonormal frame are given by 
\begin{align}
T_{(a)(b)}=T_{\mu\nu}E_{(a)}^{\mu}E_{(b)}^{\nu},
\end{align}
where $\{{E}^\mu_{(a)}\}~(a=0,1,\cdots,n-1)$ are orthonormal basis vectors satisfying 
\begin{equation}
{E}^\mu_{(a)}{E}_{(b)\mu}=\eta_{(a)(b)}=\mbox{diag}(-1,1,\cdots,1).
\end{equation}
The metric in this orthonormal frame $\eta_{(a)(b)}$ and its inverse $\eta^{(a)(b)}$ are respectively used to lower and raise the indices $(a)$ and the spacetime metric $g_{\mu\nu}$ is given by $g_{\mu\nu}=\eta_{(a)(b)}E^{(a)}_{\mu}E^{(b)}_{\nu}$.

The Hawking-Ellis classification of an energy-momentum tensor $T_{\mu\nu}$ is performed according to eigenvalues and eigenvectors of the following eigenvalue equations;
\begin{align}
T^{(a)(b)} n_{(b)}=\lambda \eta^{(a)(b)} n_{(b)}~~\Leftrightarrow~~T^{\mu\nu}n_\nu=\lambda g^{\mu\nu} n_\nu,\label{eigen-eq}
\end{align}
where $n^{(a)}=E^{(a)}_\mu n^\mu$~\cite{he1973,Maeda:2018hqu}.
The eigenvalues $\lambda$ are determined by the characteristic equation:
\begin{align}
\det \left(T^{(a)(b)}-\lambda \eta^{(a)(b)}\right)=0.\label{characteristic}
\end{align}
In arbitrary $n(\ge 3)$ dimensions that $T_{\mu\nu}$ can be classified into four types depending on the properties of its eigenvectors as summarized in Table~\ref{table:scalar+1}~\cite{he1973,Maeda:2018hqu}.
In two dimensions ($n=2$), $T_{\mu\nu}$ is classified into the type I, II, or IV.
It is noted that, in a static region of spacetime, $T_{\mu\nu}$ is of type I in any gravitation theory whose Lagrangian is a function of the Riemann tensor and metric~\cite{Maeda:2020dfp}.
\begin{table}[htb]
\begin{center}
\caption{\label{table:scalar+1} Eigenvectors of type-I--IV energy-momentum tensors.}
\scalebox{1.0}{
\begin{tabular}{|c|c|c|c|}
\hline
Type & Eigenvectors \\ \hline\hline
I & 1 timelike, $n-1$ spacelike \\ \hline
II & 1 null (doubly degenerated), $n-2$ spacelike \\ \hline
III & 1 null (triply degenerated), $n-3$ spacelike \\ \hline
IV & 2 complex, $n-2$ spacelike \\ 
\hline
\end{tabular} 
}\end{center}
\end{table}

For each type of $T_{\mu\nu}$, equivalent expressions of the standard energy conditions are available~\cite{Maeda:2018hqu}.
The standard energy conditions for $T_{\mu\nu}$ are stated as follows:
\begin{itemize}
\item {\it Null} energy condition (NEC): $T_{\mu\nu} k^\mu k^\nu\ge 0$ for any null vector $k^\mu$.
\item {\it Weak} energy condition (WEC): $T_{\mu\nu} v^\mu v^\nu\ge 0$ for any timelike vector $v^\mu$.
\item {\it Dominant} energy condition (DEC): $T_{\mu\nu} v^\mu v^\nu\ge 0$ and $J_\mu J^\mu\le 0$ hold for any timelike vector $v^\mu$, where $J^\mu:=-T^\mu_{\phantom{\mu}\nu}v^\nu$ is an energy-flux vector for an observer with its tangent vector $v^\mu$. 
\item {\it Strong} energy condition (SEC): $\left(T_{\mu\nu}-\frac{1}{n-2}Tg_{\mu\nu}\right) v^\mu v^\nu\ge 0$ for any timelike vector $v^\mu$.
\end{itemize}
The SEC is equivalent to the timelike convergence condition $R_{\mu\nu}v^{\mu}v^{\nu} \ge 0$ for any timelike vector in general relativity and defined only for $n\ge 3$.
We note that, although $n\ge 3$ is assumed in~\cite{Maeda:2018hqu}, the results presented there for the NEC, WEC, and DEC are valid also for $n=2$.

By a local Lorentz transformation ${E}^\mu_{(a)}\to{\tilde E}^\mu_{(a)}:=L_{(a)}^{~~(b)}{E}^\mu_{(b)}$ with $L_{(a)}^{~~(b)}$ satisfying $L_{(a)}^{~~(c)}L_{(b)}^{~~(d)}\eta_{(c)(d)}=\eta_{(a)(b)}$, $T^{(a)(b)}$ of each type is written in a canonical form~\cite{he1973,Maeda:2018hqu,Martin-Moruno:2017exc}.
The canonical form of type I is
\begin{equation}
\label{T-typeI}
T^{(a)(b)}=\mbox{diag}(\rho,p_1,p_2,\cdots,p_{n-1})
\end{equation}
and equivalent expressions of the standard energy conditions are given as follows:
\begin{align}
\mbox{NEC}:&~~\rho+p_i\ge 0~~\mbox{for}~~i=1,2,\cdots,n-1,\label{NEC-I}\\
\mbox{WEC}:&~~\rho\ge 0\mbox{~in addition to NEC},\label{WEC-I}\\
\mbox{DEC}:&~~\rho-p_i\ge 0~~\mbox{for}~~i=1,2,\cdots,n-1\mbox{~in addition to WEC},\label{DEC-I}\\
\mbox{SEC}:&~~(n-3)\rho+\mbox{$\sum_{j=1}^{n-1}$}p_j\ge 0~~\mbox{~in addition to NEC}\label{SEC-I}
\end{align}
The canonical form of type II is
\begin{equation} 
\label{T-typeII}
T^{(a)(b)}=\left( 
\vphantom{\begin{array}{c}1\\1\\1\\1\\1\\1\end{array}}
\begin{array}{cccccc}
\rho+\nu &\nu&0&0&\cdots &0\\
\nu&-\rho+\nu&0&0&\cdots &0\\
0&0&p_2&0&\cdots&0 \\
0&0&0&\ddots&\vdots&\vdots \\
\vdots&\vdots&\vdots&\cdots&\ddots&0\\
0&0&0 &\cdots&0&p_{n-1}
\end{array}
\right)
\end{equation}
with $\nu\ne 0$ and equivalent expressions of the standard energy conditions are 
\begin{align}
\mbox{NEC}:&~~\nu\ge 0\mbox{~and~}\rho+p_i\ge 0~~\mbox{for}~~i=2,3,\cdots,n-1,\label{NEC-II}\\
\mbox{WEC}:&~~\rho\ge 0\mbox{~in addition to NEC},\label{WEC-II}\\
\mbox{DEC}:&~~\rho-p_i\ge 0~~\mbox{for}~~i=2,3,\cdots,n-1\mbox{~in addition to WEC},\label{DEC-II}\\
\mbox{SEC}:&~~(n-4)\rho+\mbox{$\sum_{j=2}^{n-1}$}p_j\ge 0~~\mbox{~in addition to NEC}\label{SEC-II}
\end{align}
The canonical form of type III is
\begin{equation} 
\label{T-typeIII}
T^{(a)(b)}=\left( 
\vphantom{\begin{array}{c}1\\1\\1\\1\\1\\1\\1\end{array}}
\begin{array}{ccccccc}
\rho+\nu &\nu&\zeta&0&0&\cdots &0\\
\nu &-\rho+\nu&\zeta&0&0&\cdots &0\\
\zeta&\zeta&-\rho&0&0&\cdots &0\\
0&0&0&p_3&0&\cdots&0 \\
0&0&0&0&\ddots&\vdots&\vdots \\
\vdots&\vdots&\vdots&\vdots&\cdots&\ddots&0\\
0&0&0&0 &\cdots&0&p_{n-1}
\end{array}
\right)
\end{equation}
with $\zeta\ne 0$ and the type III energy-momentum tensor violates all the standard energy conditions.
In fact, one can set $\nu=0$ for type III without loss of generality by a local Lorentz transformation as presented in Ref.~\cite{Martin-Moruno:2017exc}.
Nevertheless, the present form (\ref{T-typeIII}) admitting the type-II limit $\zeta\to 0$ may be useful.
The canonical form of type IV is
\begin{equation} 
\label{T-typeIV}
T^{(a)(b)}=\left( 
\vphantom{\begin{array}{c}1\\1\\1\\1\\1\\1\end{array}}
\begin{array}{cccccc}
\rho &\nu&0&0&\cdots &0\\
\nu&-\rho&0&0&\cdots &0\\
0&0&p_2&0&\cdots&0 \\
0&0&0&\ddots&\vdots&\vdots \\
\vdots&\vdots&\vdots&\cdots&\ddots&0\\
0&0&0 &\cdots&0&p_{n-1}
\end{array}
\right)
\end{equation}
with $\nu\ne 0$ and the type IV energy-momentum tensor violates all the standard energy conditions.
Although a different form of $T^{(a)(b)}$ for type IV in the textbook~\cite{he1973}, the expression (\ref{T-typeIV}) may be more useful as pointed out in Ref.~\cite{Martin-Moruno:2017exc}.

\subsection{Main results}
\label{sec:results}

If one finds orthonormal basis one-forms ${E}_\mu^{(a)}$ in a given spacetime to provide $T^{(a)(b)}$ in canonical forms (\ref{T-typeI})--(\ref{T-typeIV}), one can immediately identify the spacetime regions where the standard energy conditions are satisfied or violated.
However, a natural set of ${E}_\mu^{(a)}$ read off from the metric do not generally provide canonical forms of $T^{(a)(b)}$.
For this reason, equivalent expressions of the energy conditions for $T^{(a)(b)}$ in a more general form must be useful.

In particular, as the simplest nontrivial case, we consider the following form of $T^{(a)(b)}$ in the present paper:
\begin{equation} 
\label{T-general}
T^{(a)(b)}=\left( 
\vphantom{\begin{array}{c}1\\1\\1\\1\\1\\1\end{array}}
\begin{array}{cccccc}
T^{(0)(0)} &T^{(0)(1)}&0&0&\cdots &0\\
T^{(0)(1)}&T^{(1)(1)}&0&0&\cdots &0\\
0&0&p_2&0&\cdots&0 \\
0&0&0&\ddots&\vdots&\vdots \\
\vdots&\vdots&\vdots&\cdots&\ddots&0\\
0&0&0 &\cdots&0&p_{n-1}
\end{array}
\right).
\end{equation}
One often encounters the above form of $T^{(a)(b)}$ with $p_2=p_3=\cdots=p_{n-1}$ in a spherically symmetric spacetime, for example.
The Hawking-Ellis type of this $T^{(a)(b)}$ is the same as the type of the two-dimensional portion $T^{(\alpha)(\beta)}~(\alpha,\beta=0,1)$ because Lorentz-covariant eigenvectors $n_{(b)}~(b=2,3,\cdots,n-1)$ corresponding to the eigenvalues $p_2, p_3, \cdots,p_{n-1}$ are all spacelike, namely $\eta^{(a)(b)}n_{(a)}n_{(b)}(=g^{\mu\nu}n_\mu n_\nu)>0$ holds.
The Hawking-Ellis type of $T^{(\alpha)(\beta)}$ is determined by the following two-dimensional eigenvalue equations:
\begin{align}
T^{(\alpha)(\beta)} n_{(\beta)}=\lambda \eta^{(\alpha)(\beta)} n_{(\beta)}.\label{eigen-eq-twodim}
\end{align}
The eigenvalues $\lambda$ are determined by the characteristic equation:
\begin{align}
\det \left(T^{(\alpha)(\beta)}-\lambda \eta^{(\alpha)(\beta)}\right)=0.\label{characteristic-twodim}
\end{align}

In fact, $T^{(\alpha)(\beta)}$ is of (i) type I if all the eigenvalues $\lambda$ determined by the characteristic equation (\ref{characteristic-twodim}) are real and non-degenerate, (ii) type II if the eigenvalues are degenerate, and (iii) type IV if the eigenvalues are complex~\cite{he1973,Maeda:2018hqu}.
Type III is not possible since the eigenvalues cannot be triply degenerated in two dimensions.
Thus, one can show the following lemma.
\begin{lm}[Hawking-Ellis type]
\label{Prop:HE-type}
The Hawking-Ellis type of the energy-momentum tensor (\ref{T-general}) is type I if $T^{(0)(1)}=0$.
If $T^{(0)(1)}\ne 0$, it is determined as
\begin{align}
&(T^{(0)(0)}+T^{(1)(1)})^2> 4(T^{(0)(1)})^2~~\Rightarrow~~\mbox{\rm Type~I},\\
&(T^{(0)(0)}+T^{(1)(1)})^2= 4(T^{(0)(1)})^2~~\Rightarrow~~\mbox{\rm Type~II},\\
&(T^{(0)(0)}+T^{(1)(1)})^2< 4(T^{(0)(1)})^2~~\Rightarrow~~\mbox{\rm Type~IV}.
\end{align}
\end{lm}
{\it Proof:}
If $T^{(0)(1)}=0$ holds, $T^{(a)(b)}$ is diagonal and hence it is of type I.
Hereafter we assume $T^{(0)(1)}\ne 0$.
For the two-dimensional part of Eq.~(\ref{T-general}), the eigenvalue equations (\ref{eigen-eq-twodim}) are written as
\begin{align}
&T^{(0)(0)} n_{(0)}+T^{(0)(1)} n_{(1)}=-\lambda n_{(0)},\\
&T^{(1)(0)} n_{(0)}+T^{(1)(1)} n_{(1)}=\lambda n_{(1)}
\end{align}
and the characteristic equation~(\ref{characteristic-twodim}) is solved to give $\lambda=\lambda_\pm$, where
\begin{align}
\lambda_\pm:=&\frac12\left\{(T^{(1)(1)}-T^{(0)(0)})\pm \sqrt{{\cal D}}\right\},\\
{\cal D}:=&(T^{(0)(0)}+T^{(1)(1)})^2-4(T^{(0)(1)})^2.
\end{align}

If ${\cal D}>0$ holds, $\lambda_\pm$ are real and not equal and then the Lorentz-covariant eigenvectors $n_{(\beta)}$ corresponding to $\lambda_\pm$ are given by 
\begin{align}
n_{(0)}=\zeta,\qquad n_{(1)}=-\frac{(T^{(1)(1)}+T^{(0)(0)})\pm\sqrt{{\cal D}}}{2T^{(0)(1)}}\zeta,
\end{align}
where $\zeta$ is an arbitrary non-vanishing constant.
By the following expression;
\begin{align}
\eta^{(\alpha)(\beta)}n_{(\alpha)}n_{(\beta)}=\frac{\sqrt{{\cal D}}\{\sqrt{{\cal D}}\pm(T^{(1)(1)}+T^{(0)(0)})\}}{2(T^{(0)(1)})^2}\zeta^2,
\end{align}
one of these eigenvectors is timelike and the other is spacelike and therefore $T^{(a)(b)}$ is of the Hawking-Ellis type I.

If ${\cal D}=0$ holds, the eigenvalues are real and degenerate such as $\lambda_+=\lambda_-=(T^{(1)(1)}-T^{(0)(0)})/2$.
In this case the Lorentz-covariant eigenvector corresponding to $\lambda_+=\lambda_-$ is
\begin{align}
n_{(0)}=\zeta,\quad n_{(1)}=-\zeta~~&\mbox{if}~~T^{(0)(1)}=\frac12(T^{(0)(0)}+T^{(1)(1)}),\\
n_{(0)}=\zeta,\quad n_{(1)}=\zeta~~~~&\mbox{if}~~T^{(0)(1)}=-\frac12(T^{(0)(0)}+T^{(1)(1)}).
\end{align}
Since $\eta^{(\alpha)(\beta)}n_{(\alpha)}n_{(\beta)}=0$ holds in both cases, these eigenvectors are null and therefore $T^{(a)(b)}$ is of the Hawking-Ellis type II.

If ${\cal D}<0$ holds, $\lambda_\pm$ are complex and conjugate to each other and their corresponding Lorentz-covariant eigenvectors are also complex.
Therefore, $T^{(a)(b)}$ is of the Hawking-Ellis type IV.
\qed

By Lemma~\ref{Prop:HE-type}, all the standard energy conditions are violated if $(T^{(0)(0)}+T^{(1)(1)})^2< 4(T^{(0)(1)})^2$ holds.
Now let us see the claim of Lemma~\ref{Prop:HE-type} in a different manner.
In order to check whether $T^{(\alpha)(\beta)}$ is diagonalizable, we perform a local Lorentz transformation in the $(0)(1)$-plane in the orthonormal frame such that
\begin{align}
\label{Thakurta-newbasis}
\begin{aligned}
&{\tilde E}^{(0)}_\mu:=\cosh\alpha E^{(0)}_\mu-\sinh\alpha E^{(1)}_\mu,\\
&{\tilde E}^{(1)}_\mu:=-\sinh\alpha E^{(0)}_\mu+\cosh\alpha E^{(1)}_\mu.
\end{aligned}
\end{align}
With a parametrization such that
\begin{align}
\cosh\alpha=\frac{1}{\sqrt{1-v^2}},\qquad \sinh\alpha=\frac{v}{\sqrt{1-v^2}},
\end{align}
where $v$ is real and satisfies $-1<v<1$, orthonormal components with the new basis one-forms are computed to give
\begin{align}
&{\tilde T}^{(0)(0)}:=T^{\mu\nu}{\tilde E}^{(0)}_\mu {\tilde E}^{(0)}_\nu=\frac{1}{1-v^2}\left(T^{(0)(0)}-2vT^{(0)(1)}+v^2T^{(1)(1)}\right),\label{G00-new}\\
&{\tilde T}^{(0)(1)}:=T^{\mu\nu}{\tilde E}^{(0)}_\mu {\tilde E}^{(1)}_\nu=\frac{1}{1-v^2}\left\{-vT^{(0)(0)}+(1+v^2)T^{(0)(1)}-vT^{(1)(1)}\right\},\\
&{\tilde T}^{(1)(1)}:=T^{\mu\nu}{\tilde E}^{(1)}_\mu {\tilde E}^{(1)}_\nu=\frac{1}{1-v^2}\left(v^2T^{(0)(0)}-2vT^{(0)(1)}+T^{(1)(1)}\right).\label{G11-new}
\end{align}
For $(T^{(0)(0)}+T^{(1)(1)})^2-4(T^{(0)(1)})^2<0$, ${\tilde T}^{(0)(1)}=0$ does not admit any real solution of $v$, so that ${T}^{(a)(b)}$ is not diagonalizable and of type IV.
For $(T^{(0)(0)}+T^{(1)(1)})^2-4(T^{(0)(1)})^2=0$, ${\tilde T}^{(0)(1)}=0$ admits real solutions $v=\pm 1$, so that ${T}^{(a)(b)}$ is not diagonalizable either and of type II.
For $(T^{(0)(0)}+T^{(1)(1)})^2-4(T^{(0)(1)})^2>0$, since $f(1)f(-1)=4(T^{(0)(1)})^2-(T^{(0)(0)}+T^{(1)(1)})^2(<0)$ holds, where 
\begin{align}
f(v):=&(1-v^2){\tilde T}^{(0)(1)}=T^{(0)(1)}v^2-(T^{(0)(0)}+T^{(1)(1)})v+T^{(0)(1)},
\end{align}
${\tilde T}^{(0)(1)}=0$ with $T^{(0)(1)}\ne 0$ admits a single real solution in the domain 
$-1<v<1$ given by
\begin{align}
v=\frac{1}{2T^{(0)(1)}}\biggl\{(T^{(0)(0)}+T^{(1)(1)})-\varepsilon\sqrt{(T^{(0)(0)}+T^{(1)(1)})^2-4(T^{(0)(1)})^2}\biggl\}.\label{v-def}
\end{align}
Here $\varepsilon=1(-1)$ is required for $T^{(0)(0)}+T^{(1)(1)}>(<)0$ in order to satisfy $v^2<1$, which is shown by 
\begin{align}
v^2=&1+\frac{(T^{(0)(0)}+T^{(1)(1)})^2-4(T^{(0)(1)})^2}{2(T^{(0)(1)})^2} \nonumber \\
&-\frac{\varepsilon(T^{(0)(0)}+T^{(1)(1)})}{2(T^{(0)(1)})^2}\sqrt{(T^{(0)(0)}+T^{(1)(1)})^2-4(T^{(0)(1)})^2}.
\end{align}
Therefore in this case, ${T}^{(a)(b)}$ is diagonalizable and of type I.

The following proposition is the main result of the present paper.

\begin{Prop}[Energy-condition criteria]
\label{Prop:EC-criteria-all}
For an energy-momentum tensor (\ref{T-general}) in an orthonormal frame, all the standard energy conditions are violated if $(T^{(0)(0)}+T^{(1)(1)})^2< 4(T^{(0)(1)})^2$ or $T^{(0)(0)}+T^{(1)(1)}<0$ is satisfied.
If $(T^{(0)(0)}+T^{(1)(1)})^2\ge 4(T^{(0)(1)})^2$ and $T^{(0)(0)}+T^{(1)(1)}\ge 0$ hold, equivalent expressions of the standard energy conditions are given by
\begin{align}
\mbox{NEC}:&~~T^{(0)(0)}-T^{(1)(1)}+2p_i+\sqrt{(T^{(0)(0)}+T^{(1)(1)})^2-4(T^{(0)(1)})^2}\ge 0 \nonumber \\
&~~\mbox{for}~~i=2,3,\cdots,n-1,\label{NEC-all}\\
\mbox{WEC}:&~~T^{(0)(0)}-T^{(1)(1)}+\sqrt{(T^{(0)(0)}+T^{(1)(1)})^2-4(T^{(0)(1)})^2}\ge 0 \nonumber \\
&\mbox{~~in addition to NEC},\label{WEC-all}\\
\mbox{DEC}:&~T^{(0)(0)}-T^{(1)(1)}-2p_i+\sqrt{(T^{(0)(0)}+T^{(1)(1)})^2-4(T^{(0)(1)})^2}\ge 0 \nonumber \\
&~~\mbox{for}~~i=2,3,\cdots,n-1~~\mbox{and}~~T^{(0)(0)}- T^{(1)(1)}\ge 0\mbox{~~in addition to WEC},\label{DEC-all}\\
\mbox{SEC}:&~~(n-4)(T^{(0)(0)}-T^{(1)(1)})+2\sum_{j=2}^{n-1}p_j\nonumber \\
&~~+(n-2)\sqrt{(T^{(0)(0)}+T^{(1)(1)})^2-4(T^{(0)(1)})^2}\ge 0\mbox{~~in addition to NEC}.\label{SEC-all}
\end{align}
\end{Prop}
{\it Proof:}
First we consider the case with $T^{(0)(1)}=0$ and identity in Eq.~(\ref{T-typeI}) as $\rho\equiv T^{(0)(0)}$ and $p_1\equiv T^{(1)(1)}$.
Then, the NEC is violated if $T^{(0)(0)}+T^{(1)(1)}<0$ holds by Eq.~(\ref{NEC-I}).
In the case of $T^{(0)(0)}+T^{(1)(1)}\ge 0$, Eqs.~(\ref{NEC-I})--(\ref{SEC-I}) become
\begin{align}
\mbox{NEC}:&~~T^{(0)(0)}+p_i\ge 0~~\mbox{for}~~i=2,3,\cdots,n-1,\\
\mbox{WEC}:&~~T^{(0)(0)}\ge 0\mbox{~in addition to NEC},\\
\mbox{DEC}:&~~T^{(0)(0)}-T^{(1)(1)}\ge 0~~\mbox{and}~~T^{(0)(0)}-p_i\ge 0~~\mbox{for}~~i=2,3,\cdots,n-1\nonumber\\
&\mbox{~~in addition to WEC},\\
\mbox{SEC}:&~~(n-3)T^{(0)(0)}+T^{(1)(1)}+\mbox{$\sum_{j=2}^{n-1}$}p_j\ge 0~~\mbox{~in addition to NEC},
\end{align}
which are identical to Eqs.~(\ref{NEC-all})--(\ref{SEC-all}) with $T^{(0)(1)}=0$ and $T^{(0)(0)}+T^{(1)(1)}\ge 0$.

Next we consider the case with $T^{(0)(1)}\ne 0$.
If $(T^{(0)(0)}+T^{(1)(1)})^2< 4(T^{(0)(1)})^2$ holds, $T^{\mu\nu}$ is of type IV by Lemma~\ref{Prop:HE-type}, so that all the standard energy conditions are violated.

If $(T^{(0)(0)}+T^{(1)(1)})^2= 4(T^{(0)(1)})^2(\ne 0)$ holds, $T^{\mu\nu}$ is of type II by Lemma~\ref{Prop:HE-type}.
In the case of $T^{(0)(1)}=(T^{(0)(0)}+T^{(1)(1)})/2$, the energy-momentum tensor (\ref{T-general}) is in the type II form~(\ref{T-typeII}) with 
\begin{align}
\nu=&\frac12(T^{(0)(0)}+T^{(1)(1)}),\qquad \rho=\frac12(T^{(0)(0)}-T^{(1)(1)}).\label{nu-rho}
\end{align}
In the case of $T^{(0)(1)}=-(T^{(0)(0)}+T^{(1)(1)})/2$, with a basis one-form ${E}^{(1)}_\mu$ replaced by $-{E}^{(1)}_\mu$, the energy-momentum tensor (\ref{T-general}) is in the type II form~(\ref{T-typeII}) with Eq.~(\ref{nu-rho}).
Thus, all the standard energy conditions are violated if $T^{(0)(0)}+T^{(1)(1)}<0$ is satisfied by Eq.~(\ref{NEC-II}) in both cases.
For $T^{(0)(0)}+T^{(1)(1)}>0$, the equivalent expressions are Eqs.~(\ref{NEC-all})--(\ref{SEC-all}) by Eqs.~(\ref{NEC-II})--(\ref{SEC-II}).

If $(T^{(0)(0)}+T^{(1)(1)})^2> 4(T^{(0)(1)})^2(\ne 0)$ holds, $T^{\mu\nu}$ is of type I by Lemma~\ref{Prop:HE-type}.
With $v$ given by Eq.~(\ref{v-def}), we obtain $\tilde{T}^{(0)(1)}=0$ and Eqs.~(\ref{G00-new}) and (\ref{G11-new}) become
\begin{align}
&{\tilde T}^{(0)(0)}=\frac12\biggl\{(T^{(0)(0)}-T^{(1)(1)})+\varepsilon\sqrt{(T^{(0)(0)}+T^{(1)(1)})^2-4(T^{(0)(1)})^2}\biggl\},\label{G00-new2}\\
&{\tilde T}^{(1)(1)}=\frac12\biggl\{-(T^{(0)(0)}-T^{(1)(1)})+\varepsilon\sqrt{(T^{(0)(0)}+T^{(1)(1)})^2-4(T^{(0)(1)})^2}\biggl\}.\label{G11-new2}
\end{align}
Identifying $\rho\equiv {\tilde T}^{(0)(0)}$, $p_1\equiv {\tilde T}^{(1)(1)}$ with Eqs.~(\ref{G00-new2}) and (\ref{G11-new2}) in Eq.~(\ref{T-typeI}), we obtain
\begin{align}
&\rho+p_1=\varepsilon\sqrt{(T^{(0)(0)}+T^{(1)(1)})^2-4(T^{(0)(1)})^2},\label{rho+p1-all}\\
&\rho+p_i=\frac12\biggl[(T^{(0)(0)}-T^{(1)(1)})+\varepsilon\sqrt{(T^{(0)(0)}+T^{(1)(1)})^2-4(T^{(0)(1)})^2}\biggl]+p_i,\\
&\rho-p_1=T^{(0)(0)}-T^{(1)(1)},\\
&\rho-p_i=\frac12\biggl[(T^{(0)(0)}-T^{(1)(1)})+\varepsilon\sqrt{(T^{(0)(0)}+T^{(1)(1)})^2-4(T^{(0)(1)})^2}\biggl]-p_i,\\
&(n-3)\rho+\sum_{j=1}^{n-1}p_j=\frac{n-4}{2}(T^{(0)(0)}-T^{(1)(1)})+\sum_{j=2}^{n-1}p_j\nonumber \\
&~~~~~~~~~~~~~~~~~~~~~~~~~~+\frac{\varepsilon(n-2)}{2}\sqrt{(T^{(0)(0)}+T^{(1)(1)})^2-4(T^{(0)(1)})^2},
\end{align}
where $i=2,3,\cdots,n-1$.
If $T^{(0)(0)}+T^{(1)(1)}<0$ is satisfied, we have $\varepsilon=-1$, so that all the standard energy conditions are violated by Eqs.~(\ref{NEC-I}) and (\ref{rho+p1-all}).
For $T^{(0)(0)}+T^{(1)(1)}>0$, we have $\varepsilon=1$ and then the equivalent expressions are Eqs.~(\ref{NEC-all})--(\ref{SEC-all}) by Eqs.~(\ref{NEC-I})--(\ref{SEC-I}).
\qed

\section{Applications}
\label{sec:applications}

In this section, as demonstrations, we apply the results obtained in the previous section to four different systems.

\subsection{Spacetime with a base manifold}
\label{sec:application1}

We consider the following warped spacetime $(M^n,g_{\mu\nu})$ of a two-dimensional Lorentzian spacetime $(M^2, g_{AB})$ and an $(n-2)$-dimensional Riemannian base manifold $(K^{n-2},\gamma_{ij})$:
\begin{align}
\D s^2 =g_{AB}(y)\D y^A\D y^B +R^2(y)\gamma_{ij}(z)\D z^i\D z^j,\label{eq:ansatz}
\end{align} 
where $y^A~(A=0,1)$ and $z^i~(i=2,3,\cdots,n-1)$ are coordinates on $(M^2, g_{AB})$ and $(K^{n-2},\gamma_{ij})$, respectively.
Here we assume that gravitational field equations give an energy-momentum tensor in the form of 
\begin{align}
T_{\mu \nu }\D x^\mu \D x^\nu =&T_{AB}(y)\D y^A\D y^B +p_{\rm t}(y)R(y)^2\gamma_{ij}(z)\D z^i\D z^j. \label{Tmunu}
\end{align} 
General relativity with or without a cosmological constant and a variety of scalar-tensor theories satisfy this condition if $(K^{n-2},\gamma_{ij})$ is an Einstein space.
In a more general Lovelock gravity~\cite{Lovelock:1971yv}, this condition is satisfied if $(K^{n-2},\gamma_{ij})$ is an Einstein space satisfying additional conditions~\cite{Dotti:2005rc,Ray:2015ava,Ohashi:2015xaa}.

One may introduce basis one-forms $\{{E}_\mu^{(a)}\}$ in the spacetime (\ref{eq:ansatz}) such that
\begin{align}
{E}_\mu^{(\alpha)}\D x^\mu={E}_A^{(\alpha)}\D y^A,\qquad {E}_\mu^{(k)}\D x^\mu=R{e}_i^{(k)}\D z^i, \label{standard-basis}
\end{align}
where basis one-forms $\{{E}_A^{(\alpha)}\}~(\alpha=0,1)$ on $(M^2, g_{AB})$ satisfy
\begin{equation}
{E}^A_{(\alpha)}{E}_{(\beta)A}=\eta_{(\alpha)(\beta)}=\mbox{diag}(-1,1)
\end{equation}
and ${e}_i^{(k)}~(k=2,3,\cdots,n-1)$ are basis one-forms on $(K^{n-2},\gamma_{ij})$ satisfying 
\begin{align}
\gamma_{ij}=\delta_{(k)(l)}e^{(k)}_{i}e^{(l)}_{j}~\leftrightarrow~\gamma^{ij}e^{(k)}_{i}e^{(l)}_{j}=\delta^{(k)(l)}.\label{e-K}
\end{align}
With this set of basis one-forms, non-zero components of $T^{(a)(b)}$ are $T^{(0)(0)}$, $T^{(0)(1)}(=T^{(1)(0)})$, $T^{(1)(1)}$, and $T^{(2)(2)}=T^{(3)(3)}=\cdots=T^{(n-1)(n-1)}=p_{\rm t}$.
Now we apply Proposition~\ref{Prop:EC-criteria-all} with three different coordinate systems on $(M^2, g_{AB})$.

\begin{Coro}[Energy conditions in diagonal coordinates]
\label{Prop:EC-diagonal}
For an energy-momentum tensor (\ref{Tmunu}) in the spacetime (\ref{eq:ansatz}) in the following diagonal coordinates $y^A=(t,x)$ on $(M^2,g_{AB})$;
\begin{align}
\D s^2=-e^{2\Phi(t,x)}\D t^2+e^{2\Psi(t,x)}\D x^2+R(t,x)^2\gamma_{ij}(z)\D z^i\D z^j,\label{diagonal-c}
\end{align}
all the energy conditions are violated in a region with $e^{-2\Phi}T_{tt}+e^{-2\Psi}T_{xx}<0$ or ${\cal D}_1<0$, where
\begin{align}
{\cal D}_1:=(e^{-2\Phi}T_{tt}+e^{-2\Psi}T_{xx}+2e^{-\Phi-\Psi}T_{tx})(e^{-2\Phi}T_{tt}+e^{-2\Psi}T_{xx}-2e^{-\Phi-\Psi}T_{tx}). \label{def-D1}
\end{align}
In a region where ${\cal D}_1\ge 0$ and $e^{-2\Phi}T_{tt}+e^{-2\Psi}T_{xx}\ge 0$ hold, equivalent expressions of the standard energy conditions are given by
\begin{align}
\mbox{NEC}:&~~e^{-2\Phi}T_{tt}-e^{-2\Psi}T_{xx}+2p_{\rm t}+\sqrt{{\cal D}_1}\ge 0,\label{NEC-diag}\\
\mbox{WEC}:&~~e^{-2\Phi}T_{tt}-e^{-2\Psi}T_{xx}+\sqrt{{\cal D}_1}\ge 0\mbox{~~in addition to NEC},\label{WEC-diag}\\
\mbox{DEC}:&~e^{-2\Phi}T_{tt}-e^{-2\Psi}T_{xx}-2p_{\rm t}+\sqrt{{\cal D}_1}\ge 0 \nonumber \\
&\mbox{and}~~e^{-2\Phi}T_{tt}-e^{-2\Psi}T_{xx}\ge 0\mbox{~~in addition to WEC},\label{DEC-diag}\\
\mbox{SEC}:&~~(n-4)(e^{-2\Phi}T_{tt}-e^{-2\Psi}T_{xx})+2(n-2)p_{\rm t}+(n-2)\sqrt{{\cal D}_1}\ge 0\nonumber\\
&\mbox{~~in addition to NEC}.\label{SEC-diag}
\end{align}
\end{Coro}
{\it Proof:}
With the following orthonormal basis one-forms on $(M^2,g_{AB})$
\begin{align}
{E}_\mu^{(0)}\D x^\mu=&-e^{\Phi}\D t,\qquad {E}_\mu^{(1)}\D x^\mu=e^{\Psi}\D x,
\end{align}
we obtain
\begin{align}
T^{(0)(0)}=&e^{-2\Phi}T_{tt},\qquad T^{(1)(1)}=e^{-2\Psi}T_{xx},\qquad T^{(0)(1)}=e^{-\Phi-\Psi}T_{tx},
\end{align}
which give
\begin{align}
&(T^{(0)(0)}+T^{(1)(1)})^2- 4(T^{(0)(1)})^2={\cal D}_1,
\end{align}
where ${\cal D}_1$ is defined by Eq.~(\ref{def-D1}).
Then, the corollary follows from Proposition~\ref{Prop:EC-criteria-all}.
\qed

\begin{Coro}[Energy conditions in single-null coordinates]
\label{Prop:EC-single}
For an energy-momentum tensor (\ref{Tmunu}) in the spacetime (\ref{eq:ansatz}) in the following single-null coordinates $y^A=(u,r)$ on $(M^2,g_{AB})$;
\begin{align}
\D s^2=-f(u,r)\D u^2-2\epsilon e^{-\delta(u,r)}\D u\D r+R(u,r)^2\gamma_{ij}(z)\D z^i\D z^j,\label{single-null-c}
\end{align}
where $\epsilon=\pm 1$, all the energy conditions are violated in a region where ${\cal D}_2<0$ or $T_{uu}-\epsilon fe^{\delta}T_{ur}+(f^2/4+1)e^{2\delta}T_{rr}<0$ is satisfied, where
\begin{align}
{\cal D}_2:=e^{2\delta}T_{rr}(4T_{uu}-4\epsilon fe^{\delta}T_{ur}+f^2e^{2\delta}T_{rr}). \label{def-D2}
\end{align}
In a region where ${\cal D}_2\ge 0$ and $T_{uu}-\epsilon fe^{\delta}T_{ur}+(f^2/4+1)e^{2\delta}T_{rr}\ge 0$ hold, equivalent expressions of the standard energy conditions are given by
\begin{align}
\mbox{NEC}:&~~2\epsilon e^{\delta}T_{ur}-fe^{2\delta}T_{rr}+2p_{\rm t}+\sqrt{{\cal D}_2}\ge 0,\label{NEC-single}\\
\mbox{WEC}:&~~2\epsilon e^{\delta}T_{ur}-fe^{2\delta}T_{rr}+\sqrt{{\cal D}_2}\ge 0\mbox{~~in addition to NEC},\label{WEC-single}\\
\mbox{DEC}:&~~2\epsilon e^{\delta}T_{ur}-fe^{2\delta}T_{rr}-2p_{\rm t}+\sqrt{{\cal D}_2}\ge 0 \nonumber \\
&~~\mbox{and}~~2\epsilon e^{\delta}T_{ur}-fe^{2\delta}T_{rr}\ge 0\mbox{~~in addition to WEC},\label{DEC-single}\\
\mbox{SEC}:&~~(n-4)(2\epsilon e^{\delta}T_{ur}-fe^{2\delta}T_{rr})+2(n-2)p_{\rm t}+(n-2)\sqrt{{\cal D}_2}\ge 0 \nonumber \\
&\mbox{~~in addition to NEC}.\label{SEC-single}
\end{align}
\end{Coro}
{\it Proof:}
With the following orthonormal basis one-forms on $(M^2,g_{AB})$
\begin{align}
{E}_\mu^{(0)}\D x^\mu=&-\frac{1}{\sqrt{2}}\left(1+\frac{f}{2}\right)\D u-\frac{\epsilon}{\sqrt{2}}e^{-\delta}\D r,\\
{E}_\mu^{(1)}\D x^\mu=&-\frac{1}{\sqrt{2}}\left(1-\frac{f}{2}\right)\D u+\frac{\epsilon}{\sqrt{2}}e^{-\delta}\D r,
\end{align}
we obtain
\begin{align}
T^{(0)(0)}=&\frac18\left\{4T_{uu}-4\epsilon(f-2)e^{\delta}T_{ur}+(f-2)^2e^{2\delta}T_{rr}\right\},\\
T^{(1)(1)}=&\frac18\left\{4T_{uu}-4\epsilon(f+2)e^{\delta}T_{ur}+(f+2)^2e^{2\delta}T_{rr}\right\},\\
T^{(0)(1)}=&\frac18\left\{-4T_{uu}+4\epsilon fe^{\delta}T_{ur}-(f+2)(f-2)e^{2\delta}T_{rr}\right\},
\end{align}
which give
\begin{align}
&(T^{(0)(0)}+T^{(1)(1)})^2- 4(T^{(0)(1)})^2={\cal D}_2,\\
&T^{(0)(0)}+T^{(1)(1)}=T_{uu}-\epsilon fe^{\delta}T_{ur}+\biggl(\frac14f^2+1\biggl)e^{2\delta}T_{rr},\\
&T^{(0)(0)}-T^{(1)(1)}=2\epsilon e^{\delta}T_{ur}-fe^{2\delta}T_{rr}.
\end{align}
where ${\cal D}_2$ is defined by Eq.~(\ref{def-D2}).
Then, the corollary follows from Proposition~\ref{Prop:EC-criteria-all}.
\qed

\begin{Coro}[Energy conditions in double-null coordinates]
\label{Prop:EC-double}
For an energy-momentum tensor (\ref{Tmunu}) in the spacetime (\ref{eq:ansatz}) in the following double-null coordinates $y^A=(u,v)$ on $(M^2,g_{AB})$;
\begin{align}
\D s^2=-2e^{-f(u,v)}\D u\D v+R(u,v)^2\gamma_{ij}(z)\D z^i\D z^j,\label{double-null-c}
\end{align}
all the standard energy conditions are violated in a region where $T_{uu}T_{vv}<0$ or $T_{uu}+T_{vv}<0$ is satisfied.
In a region where $T_{uu}T_{vv}\ge 0$ and $T_{uu}+T_{vv}\ge 0$ hold, equivalent expressions of the standard energy conditions are given by
\begin{align}
\mbox{NEC}:&~~T_{uv}+p_{\rm t}e^{-f}+\sqrt{T_{uu}T_{vv}}\ge 0,\label{NEC-double}\\
\mbox{WEC}:&~~T_{uv}+\sqrt{T_{uu}T_{vv}}\ge 0~~\mbox{~in addition to NEC},\label{WEC-double}\\
\mbox{DEC}:&~T_{uv}-p_{\rm t}e^{-f}+\sqrt{T_{uu}T_{vv}}\ge 0~~\mbox{and}~~T_{uv}\ge 0\mbox{~in addition to WEC},\label{DEC-double}\\
\mbox{SEC}:&~~(n-4)T_{uv}+(n-2)p_{\rm t}e^{-f}+(n-2)\sqrt{T_{uu}T_{vv}}\ge 0\mbox{~~in addition to NEC}.\label{SEC-double}
\end{align}
\end{Coro}
{\it Proof:}
With the following orthonormal basis one-forms on $(M^2,g_{AB})$
\begin{align}
&{E}_\mu^{(0)}\D x^\mu=-\frac{1}{\sqrt{2}}e^{-f/2}(\D u+\D v)\\
&{E}_\mu^{(1)}\D x^\mu=\frac{1}{\sqrt{2}}e^{-f/2}(\D u-\D v).\label{E1-double}
\end{align}
we obtain
\begin{align}
T^{(0)(0)}=&\frac12e^f(T_{uu}+2T_{uv}+T_{vv}),\\
T^{(1)(1)}=&\frac12e^f(T_{uu}-2T_{uv}+T_{vv}),\\
T^{(0)(1)}=&\frac12e^f(T_{uu}-T_{vv}),
\end{align}
which give
\begin{align}
&(T^{(0)(0)}+T^{(1)(1)})^2- 4(T^{(0)(1)})^2=4e^{2f}T_{uu}T_{vv},\\
&T^{(0)(0)}+T^{(1)(1)}=e^f(T_{uu}+T_{vv}),\\
&T^{(0)(0)}-T^{(1)(1)}=2e^f T_{uv}.
\end{align}
Then, the corollary follows from Proposition~\ref{Prop:EC-criteria-all}.
\qed

Corollary~\ref{Prop:EC-double} shows that the NEC implies $T_{uu}\ge 0$ and $T_{vv}\ge 0$ and the DEC implies $T_{uu}\ge 0$, $T_{vv}\ge 0$, and $T_{uv}\ge 0$ in the spacetime with a metric given by Eq.~(\ref{double-null-c}).

\subsection{Generalized Kerr spacetime}
\label{sec:application3}

The second application is to a stationary and axisymmetric spacetime.
In particular, we consider the following G\"urses-G\"ursey spacetime in the Boyer-Lindquist coordinates $(t,r,\theta,\varphi)$;
\begin{align}
\label{BL}
\begin{aligned}
\D s^2=&-\biggl(1-\frac{2M(r)r}{\Sigma(r,\theta)}\biggl)\D t^2-\frac{4aM(r)r\sin^2\theta}{\Sigma(r,\theta)}\D t\D\phi \\
&+\frac{\Sigma(r,\theta)}{\Delta(r)}\D r^2+\Sigma(r,\theta)\D\theta^2+\biggl(r^2+a^2+\frac{2a^2M(r)r\sin^2\theta}{\Sigma(r,\theta)}\biggl)\sin^2\theta\D\phi^2, \\
&\Sigma(r,\theta):=r^2+a^2\cos^2\theta,\qquad \Delta(r):=r^2+a^2-2rM(r),
\end{aligned}
\end{align} 
in which $a$ is a constant and $M(r)$ is a function of $r$~\cite{Gurses:1975vu}.
If $M(r)$ is constant, the metric (\ref{BL}) reduces to Kerr in the Boyer-Lindquist coordinates.
A regular null hypersurface $r=r_{\rm h}$ determined by $\Delta(r_{\rm h})=0$ is a Killing horizon associated with a Killing vector $\xi^\mu=(1,0,0,a/(r_{\rm h}^2+a^2))$.
The G\"urses-G\"ursey metric has been used to construct models of a rotating non-singular black hole~\cite{Maeda:2021jdc,Burinskii:2001bq,Toshmatov:2014nya,Azreg-Ainou:2014pra,Ghosh:2014pba,Amir:2016cen,Torres:2016pgk,Toshmatov:2017zpr,Ghosh:2020ece}.
G\"urses and G\"ursey showed that the corresponding matter field in general relativity can be interpreted as an anisotropic fluid, which is of the Hawking-Ellis type I~\cite{Gurses:1975vu}.

In the coordinates (\ref{BL}), a Killing horizon $r=r_{\rm h}$ is a coordinate singularity.
For this reason, we will study the G\"urses-G\"ursey spacetime in the following Doran coordinates $(\eta,r,\theta,\varphi)$~\cite{Doran:1999gb,Visser:2007fj,Maeda:2021jdc}:
\begin{align}
\D s^2=&-\D \eta^2+\Sigma(r,\theta)\D \theta^2+(r^2+a^2)\sin^2\theta\D\varphi^2\nonumber \\
&+\frac{\Sigma(r,\theta)}{r^2+a^2}\biggl\{\D r+\frac{\sqrt{2M(r)r(r^2+a^2)}}{\Sigma(r,\theta)}(\D \eta-a\sin^2\theta\D\varphi)\biggl\}^2,\label{Doran-g}
\end{align}
which is obtained from Eq.~(\ref{BL}) by the following coordinate transformations
\begin{align}
&\D {t}=\D\eta-\frac{\sqrt{2M(r)r(r^2+a^2)}}{\Delta(r)}\D r,\\
&\D{\phi}=\D\varphi-\frac{a}{\Delta(r)}\sqrt{\frac{2M(r)r}{r^2+a^2}}\D r.
\end{align}
Different from the Boyer-Lindquist coordinates, Killing horizons are not coordinate singularities in the Doran coordinates.

Natural orthonormal basis one-forms in the spacetime (\ref{Doran-g}) are
\begin{align}
\label{basis-rotating}
\begin{aligned}
&E^{(0)}_\mu\D x^\mu = -\D\eta,\\
&E^{(1)}_\mu\D x^\mu = \sqrt{\frac{\Sigma(r,\theta)}{r^2+a^2}}\biggl\{\D r+\frac{\sqrt{2M(r)r(r^2+a^2)}}{\Sigma(r,\theta)}(\D \eta-a\sin^2\theta\D\varphi)\biggl\},\\
&E^{(2)}_\mu\D x^\mu=\sqrt{\Sigma}\D \theta,\qquad E^{(3)}_\mu\D x^\mu=\sqrt{r^2+a^2}\sin\theta\D\varphi.
\end{aligned}
\end{align}
with which non-zero orthonormal components of the Einstein tensor $G^{(a)(b)}$ are
\begin{align}
G^{(0)(0)}=&\Sigma^{-3}[-rM''\Sigma a^2\sin^2\theta+2M'(r^4+a^2r^2-a^4\sin^2\theta\cos^2\theta)],\\
G^{(0)(3)}=&a\sin\theta\sqrt{r^2+a^2}\Sigma^{-3}[rM''\Sigma -2M'(r^2-a^2\cos^2\theta)],\\
G^{(1)(1)}=&-2r^2\Sigma^{-2}M',\\
G^{(2)(2)}=&-\Sigma^{-2}(rM''\Sigma +2M'a^2\cos^2\theta),\\
G^{(3)(3)}=&-\Sigma^{-3}[rM''\Sigma (r^2+a^2)+2a^2M'\{(r^2+a^2)\cos^2\theta-r^2\sin^2\theta\}],
\end{align}
where a prime denotes differentiation with respect to $r$.
Thus, one can use Lemma~\ref{Prop:HE-type} and Proposition~\ref{Prop:EC-criteria-all} by exchanging the index $(1)$ for $(3)$.

\begin{Coro}[Energy conditions in the G\"urses-G\"ursey spacetime]
\label{Prop:EC-GG}
For the corresponding energy-momentum tensor in the G\"urses-G\"ursey spacetime (\ref{Doran-g}) in general relativity, equivalent expressions of the standard energy conditions are given by
\begin{align}
\mbox{NEC}:&~~2M'(r^2-a^2\cos^2\theta)-rM''\Sigma\ge 0,\label{NEC-GG}\\
\mbox{WEC}:&~~M'\ge 0\mbox{~~in addition to NEC},\label{WEC-GG}\\
\mbox{DEC}:&~~rM''+2M'\ge 0\mbox{~~in addition to WEC},\label{DEC-GG}\\
\mbox{SEC}:&~~rM''\Sigma+2M'a^2\cos^2\theta\le 0\mbox{~~in addition to NEC}.\label{SEC-GG}
\end{align}
\end{Coro}
{\it Proof:}
We compute
\begin{align}
(G^{(0)(0)}+G^{(3)(3)})^2-4(G^{(0)(3)})^2=&\Sigma^{-4}[rM''\Sigma -2M'(r^2-a^2\cos^2\theta)]^2=:{\cal D}_3
\end{align}
and ${\cal D}_3=0$ implies $G^{(0)(3)}=0$, so that the corresponding energy-momentum tensor $T_{\mu\nu}$ in general relativity is of type I by Lemma~\ref{Prop:HE-type}.
By Proposition~\ref{Prop:EC-criteria-all} with
\begin{align}
&G^{(0)(0)}+G^{(3)(3)}=-\Sigma^{-3}(r^2+a^2+a^2\sin^2\theta)[rM''\Sigma -2M'(r^2-a^2\cos^2\theta)],
\end{align}
all the standard energy conditions are violated if $rM''\Sigma -2M'(r^2-a^2\cos^2\theta)>0$ holds.
If $rM''\Sigma -2M'(r^2-a^2\cos^2\theta)\le 0$ holds, we obtain
\begin{align}
\sqrt{{\cal D}_3}=-\Sigma^{-2}[rM''\Sigma -2M'(r^2-a^2\cos^2\theta)]
\end{align}
and hence
\begin{align}
&G^{(0)(0)}-G^{(3)(3)}+2G^{(1)(1)}+\sqrt{{\cal D}_3}=0,\\
&G^{(0)(0)}-G^{(3)(3)}+2G^{(2)(2)}+\sqrt{{\cal D}_3}=-2\Sigma^{-2}[rM''\Sigma -2M'(r^2-a^2\cos^2\theta)],\\
&G^{(0)(0)}-G^{(3)(3)}+\sqrt{{\cal D}_3}=4r^2M'\Sigma^{-2},\\
&G^{(0)(0)}-G^{(3)(3)}-2G^{(1)(1)}+\sqrt{{\cal D}_3}=8r^2M'\Sigma^{-2},\\
&G^{(0)(0)}-G^{(3)(3)}-2G^{(2)(2)}+\sqrt{{\cal D}_3}=2\Sigma^{-1}(rM''+2M'),\\
&G^{(1)(1)}+G^{(2)(2)}+\sqrt{{\cal D}_3}=-2\Sigma^{-2}(rM''\Sigma+2M'a^2\cos^2\theta).
\end{align}
Then, the corollary follows from Proposition~\ref{Prop:EC-criteria-all} with $n=4$.
\qed

As an alternative and more direct approach to prove Corollary~\ref{Prop:EC-GG}~\cite{Maeda:2021jdc}, we may introduce a new set of the basis one-forms $\{{\tilde E}_\mu^{(0)},{E}_\mu^{(1)},{E}_\mu^{(2)},{\tilde E}_\mu^{(3)}\}$ obtained by a local Lorentz transformation on the plane spanned by $E_\mu^{(0)}$ and $E_\mu^{(3)}$ such that 
\begin{align}
\begin{aligned}
&{\tilde E}^{(0)}_\mu=\cosh\alpha E^{(0)}_\mu-\sinh\alpha E^{(3)}_\mu,\\
&{\tilde E}^{(3)}_\mu=-\sinh\alpha E^{(0)}_\mu+\cosh\alpha E^{(3)}_\mu
\end{aligned}
\end{align}
with
\begin{align}
\cosh\alpha=\sqrt{\frac{r^2+a^2}{\Sigma(r,\theta)}},\qquad \sinh\alpha=-\frac{a\sin\theta}{\sqrt{\Sigma(r,\theta)}}.
\end{align}
Non-zero components of ${\tilde G}^{(a)(b)}=G^{\mu\nu}{\tilde E}_\mu^{(a)}{\tilde E}_\nu^{(b)}$ with the new set of basis one-forms are then
\begin{align}
{\tilde G}^{(0)(0)}=&\frac{2r^2M'}{\Sigma^2},\qquad {\tilde G}^{(1)(1)}=-\frac{2r^2M'}{\Sigma^2},\\
{\tilde G}^{(2)(2)}=&{\tilde G}^{(3)(3)}=-\frac{rM''\Sigma +2a^2 M'\cos^2\theta}{{\Sigma^2}}.
\end{align}
Therefore, the corresponding energy-momentum tensor $T^{\mu\nu}:=G^{\mu\nu}/(8\pi G)$ in general relativity is of the Hawking-Ellis type I and its orthonormal components are given in the type-I form~(\ref{T-typeI}) with 
\begin{align}
\rho=&-p_1=\frac{r^2M'}{4\pi G\Sigma^2},\qquad p_2=p_3=-\frac{rM''\Sigma+2M'a^2\cos^2\theta}{8\pi G\Sigma^2}, \label{matter-rot}
\end{align}
where $G$ is the gravitational constant.
Equation~(\ref{matter-rot}) gives
\begin{align}
\label{matter-rotating}
\begin{aligned}
&\rho+p_1=0,\qquad \rho-p_1=\frac{r^2M'}{2\pi G\Sigma^2},\\
&\rho+p_2=\rho+p_3=\frac{2M'(r^2-a^2\cos^2\theta)-rM''\Sigma}{8\pi G\Sigma^2},\\
&\rho-p_2=\rho-p_3=\frac{rM''+2M'}{8\pi G\Sigma},\\
&\rho+p_1+p_2+p_3=-\frac{rM''\Sigma+2M'a^2\cos^2\theta}{4\pi G\Sigma^2}
\end{aligned}
\end{align}
and then Corollary~\ref{Prop:EC-GG} follows from Eqs.~(\ref{NEC-I})--(\ref{SEC-I}) with Eqs.~(\ref{matter-rot}) and (\ref{matter-rotating}).

\subsection{Imperfect fluid without shear viscosity}
\label{sec:application2}

As the third application of Proposition~\ref{Prop:EC-criteria-all}, we study the energy conditions without assuming any spacetime symmetry for a relativistic imperfect fluid without shear viscosity.
The energy-momentum tensor for a relativistic imperfect fluid in $n(\ge 2)$ dimensions is given by 
\begin{align}
T_{\mu\nu}=&\rho u_\mu u_\nu+p h_{\mu\nu}+(u_\mu q_\nu+q_\mu u_\nu)+\pi_{\mu\nu},\label{fluid}
\end{align}
where $\rho$ is the energy density, $p$ is a pressure, $u^\mu$ is an $n$-velocity of the fluid element satisfying $u_\mu u^\mu=-1$, $h_{\mu\nu}:=g_{\mu\nu}+u_\mu u_\nu$ is a projection tensor satisfying $h_{\mu\nu}u^\nu=0$, $q^\mu$ is the (spacelike) heat flux vector satisfying $u_\mu q^\mu=0$, and $\pi_{\mu\nu}(=\pi_{(\mu\nu)})$ is a viscous shear tensor satisfying $\pi_{\mu\nu}u^\mu=0$.
The viscous shear tensor $\pi_{\mu\nu}$ is written as
\begin{align}
\pi_{\mu\nu}=&-\zeta\theta h_{\mu\nu}-2\eta\sigma_{\mu\nu},\label{fluid2}
\end{align}
where $\zeta(\ge 0)$ is the coefficient of bulk viscosity, $\eta(\ge 0)$ is the coefficient of shear viscosity, and the expansion $\theta$ and the shear tensor $\sigma_{\mu\nu}(=\sigma_{(\mu\nu)})$ of the fluid are defined by 
\begin{align}
&\theta:=\nabla_\mu u^\mu,\label{def-expansion}\\
&\sigma_{\mu\nu}:=\nabla_{(\mu}u_{\nu)}+a_{(\mu}u_{\nu)}-\frac{1}{n-1}\theta h_{\mu\nu}.\label{def-shear}
\end{align}
Here $a^\mu:=u^\nu \nabla_\nu u^\mu$ is an acceleration vector.
Since $a_\mu u^\mu=0$ holds, $\sigma_{\mu\nu}$ is trace-free ($\sigma^\mu_{~\mu}=0$) and satisfies $\sigma_{\mu\nu}u^\nu=0$, so that $\sigma_{\mu\nu}\equiv 0$ holds for $n=2$.

The following corollary is an $n(\ge 2)$-dimensional generalization of Corollary 1 in~\cite{kst1988} including the NEC.
\begin{Coro}[Energy conditions for imperfect fluid without shear viscosity]
\label{Prop:EC-fluid}
For an imperfect fluid given by Eqs.~(\ref{fluid})--(\ref{def-shear}) with $\eta=0$, all the standard energy conditions are violated in a region where $(\rho+p-\zeta\theta)^2< 4Q^2$ or $\rho+p-\zeta\theta<0$ is satisfied, where $Q^2:=q_\mu q^\mu$.
In a region where $(\rho+p-\zeta\theta)^2\ge 4Q^2$ and $\rho+p-\zeta\theta\ge 0$ hold, the NEC is satisfied and equivalent expressions of other energy conditions are given by
\begin{align}
\mbox{WEC}:&~~\rho-p+\zeta\theta+\sqrt{(\rho+p-\zeta\theta)^2-4Q^2}\ge 0,\label{WEC-fluid}\\
\mbox{DEC}:&~~\rho-3(p-\zeta\theta)+\sqrt{(\rho+p-\zeta\theta)^2-4Q^2}\ge 0 \nonumber \\
&~~\mbox{and}~~\rho-p+\zeta\theta\ge 0\mbox{~~in addition to WEC},\label{DEC-fluid}\\
\mbox{SEC}:&~~(n-4)\rho+n(p-\zeta\theta)+(n-2)\sqrt{(\rho+p-\zeta\theta)^2-4Q^2}\ge 0.\label{SEC-fluid}
\end{align}
\end{Coro}
{\it Proof:}
We define a scalar function $Q$ by $q_\rho=Q{\tilde q}_\mu$ and ${\tilde q}_\mu{\tilde q}^\mu=1$.
With a set of orthonormal basis vectors $\{{E}^\mu_{(a)}\}~(a=0,1,\cdots,n-1)$ with ${E}^\mu_{(0)}=u^\mu$ and ${E}^\mu_{(1)}={\tilde q}^\mu$, we obtain
\begin{align}
T_{(0)(0)}=&\rho,\qquad T_{(0)(1)}=-Q,\qquad T_{(I)(J)}=(p-\zeta\theta)\delta_{(I)(J)},
\end{align}
where $I=1,2,\cdots,n-1$.
Then, the corollary follows from Proposition~\ref{Prop:EC-criteria-all}.
\qed

In fact, it is not straightforward to generalize Corollary~\ref{Prop:EC-fluid} to the case with shear viscosity ($\eta\ne 0$) in the most general case.
Nevertheless, Lemma~\ref{Prop:HE-type} and Proposition~\ref{Prop:EC-criteria-all} can be directly applied to the case with spherical symmetry~\cite{Brassel:2021qxf}.

\subsection{Minimally coupled scalar field}
\label{sec:application4}

As the last application, without assuming any spacetime symmetry, we study the energy conditions for a minimally coupled scalar field $\phi$ with an arbitrary self-interacting potential $V(\phi)$, of which Lagrangian density is given by 
\begin{align}
{\cal L}_{\rm m}=-\biggl(\frac12\varepsilon (\nabla\phi)^2+V(\phi)\biggl),\label{Lm-scalar}
\end{align}
where $(\nabla\phi)^2:=(\nabla_\rho\phi)(\nabla^\rho\phi)$ and the parameter $\varepsilon$ is either 1 (for a real
scalar field) or $-1$ (for a ghost scalar field).
The energy-momentum tensor for $\phi$ is given by
\begin{align}
T_{\mu\nu}=\varepsilon (\nabla_\mu \phi)(\nabla_\nu \phi)-g_{\mu\nu}\biggl(\frac12 \varepsilon (\nabla\phi)^2+V(\phi)\biggl). \label{T-scalar}
\end{align}
It was shown in Proposition 20 in Ref.~\cite{Maeda:2018hqu} that the energy-momentum tensor~(\ref{T-scalar}) respects (violates) the NEC for $\varepsilon=1$ ($\varepsilon=-1$).
However, in the case of $\varepsilon=1$, only sufficient conditions were derived for other energy conditions to be respected.
Here we present a necessary and sufficient condition.
\begin{Coro}[Energy conditions for minimally coupled scalar field]
\label{Pro:EC-scalar}
If $\nabla_{\mu}\phi$ is vanishing or non-null, the energy-momentum tensor (\ref{T-scalar}) is of the Hawking-Ellis type I, while it is of type II if $\nabla_{\mu}\phi$ is non-vanishing and null.
If $\nabla_{\mu}\phi=0$ holds, the NEC is satisfied, while equivalent expressions to the WEC, DEC and SEC are $V\ge 0$, $V\ge 0$, and $V\le 0$, respectively, regardless of the sign of $\varepsilon$. 
If $\nabla_{\mu}\phi$ is non-vanishing with $\varepsilon=-1$, all the standard energy conditions are violated. 
If $\nabla_{\mu}\phi$ is non-vanishing with $\varepsilon=1$, the NEC is satisfied and equivalent expressions to the WEC, DEC and SEC depending on the signature of $\nabla_{\mu}\phi$ are as shown in the following table.
\begin{center}
\begin{tabular}{|c||c|c|c|c|c|}
\hline \hline
$\nabla_\mu\phi(\ne 0)$ with $\varepsilon=1$ & WEC & DEC & SEC \\\hline
Timelike & $V\ge (\nabla\phi)^2/2$ & $V\ge 0$ & $V\le -(n-2)(\nabla\phi)^2/2$ \\ \hline
Spacelike & $V\ge-(\nabla\phi)^2/2$ & $V\ge 0$ & $V\le 0$ \\ \hline
Null & $V\ge 0$ & $V\ge 0$ & $V\le 0$ \\ 
\hline \hline
\end{tabular} 
\end{center} 
\end{Coro}
{\it Proof}. 
Orthonormal components of $T_{\mu\nu}$ are written as
\begin{align}
T_{(a)(b)}=&T_{\mu\nu}E^\mu_{(a)}E^\nu_{(b)}=\varepsilon E^\mu_{(a)}(\nabla_\mu \phi)E^\nu_{(b)}(\nabla_\nu \phi)-\eta_{(a)(b)}\biggl(\frac12 \varepsilon (\nabla\phi)^2+V(\phi)\biggl).
\end{align}
If $\nabla_\mu\phi$ is vanishing, we obtain $T_{(a)(b)}=-\eta_{(a)(b)}V$, which is of the Hawking-Ellis type I and equivalent to a cosmological constant $\Lambda(=V)$.
Then, by Eqs.~(\ref{T-typeI})--(\ref{SEC-I}), the NEC is satisfied, the WEC and DEC are equivalent to $V\ge 0$, and the SEC is equivalent to $V\le 0$.

Hereafter we assume that $\nabla_\mu\phi$ is non-vanishing.
Then, depending on the sign of $(\nabla\phi)^2$, one may set basis one-forms at each spacetime point such that 
\begin{eqnarray}
\nabla_\mu\phi = 
\begin{cases}
\chi{E}_\mu^{(0)} & (\mbox{for~timelike~} \nabla_\mu\phi),\\
\chi{E}_\mu^{(1)} & (\mbox{for~spacelike~} \nabla_\mu\phi),\\
\chi({E}_\mu^{(0)}-{E}_\mu^{(1)}) & (\mbox{for~null~} \nabla_\mu\phi)
\end{cases}
\end{eqnarray}
without loss of generality by a local Lorentz transformation, where $\chi$ is a non-vanishing constant.
If $\nabla_\mu\phi$ is timelike, we obtain $(\nabla\phi)^2=-\chi^2$ and 
\begin{align}
\begin{aligned}
T_{(0)(0)}=&\frac12 \varepsilon \chi^2+V,\qquad T_{(1)(1)}=\frac12 \varepsilon \chi^2-V,\\
T_{(0)(1)}=&0,\qquad T_{(i)(j)}=\delta_{(i)(j)}\biggl(\frac12 \varepsilon \chi^2-V\biggl).
\end{aligned}
\end{align}
If $\nabla_\mu\phi$ is spacelike, we obtain $(\nabla\phi)^2=\chi^2$ and 
\begin{align}
\begin{aligned}
T_{(0)(0)}=&\frac12 \varepsilon \chi^2+V,\qquad T_{(1)(1)}=\frac12 \varepsilon \chi^2-V,\\
T_{(0)(1)}=&0,\qquad T_{(i)(j)}=-\delta_{(i)(j)}\biggl(\frac12 \varepsilon \chi^2+V\biggl).
\end{aligned}
\end{align}
If $\nabla_\mu\phi$ is null, we obtain $(\nabla\phi)^2=0$ and 
\begin{align}
\begin{aligned}
T_{(0)(0)}=&\varepsilon \chi^2+V,\qquad T_{(1)(1)}=\varepsilon \chi^2-V,\\
T_{(0)(1)}=&\varepsilon \chi^2,\qquad T_{(i)(j)}=-\delta_{(i)(j)}V.
\end{aligned}
\end{align}
Then, the corollary follows from Lemma~\ref{Prop:HE-type} and Proposition~\ref{Prop:EC-criteria-all}.
\qed

Corollary~\ref{Pro:EC-scalar} shows that a massless scalar field ($V\equiv 0$) with $\varepsilon=1$ satisfies all the standard energy conditions.

\section{Summary}
\label{sec:summary}

In the present paper, we have derived criteria for the Hawking-Ellis types and the standard energy conditions for the case where the energy-momentum tensor $T_{\mu\nu}$ has a single off-diagonal ``space-time'' component in an orthonormal frame.
In Sec.~\ref{sec:main}, we have shown that, for an energy-momentum tensor of which components in an orthonormal frame is given by Eq.~(\ref{T-general}), its Hawking-Ellis type is identified by Lemma~\ref{Prop:HE-type} and one can check the standard energy conditions by Proposition~\ref{Prop:EC-criteria-all}.
In Sec.~\ref{sec:applications}, we have adopted those results to four different systems.

In Sec.~\ref{sec:application1}, we have derived equivalent expressions to the energy conditions for an energy-momentum tensor (\ref{Tmunu}) in terms of the coordinate components in the spacetime (\ref{eq:ansatz}) in three different coordinate systems.
In Sec.~\ref{sec:application3}, we have applied our results to a stationary and axisymmetric G\"urses-G\"ursey spacetime (\ref{Doran-g}) in the Doran coordinates and derived equivalent expressions to the energy conditions in general relativity.
In Secs.~\ref{sec:application2} and \ref{sec:application4}, without assuming any spacetime symmetry, equivalent expressions to the energy conditions have been derived for an imperfect fluid without shear viscosity and a minimally coupled scalar field, respectively.

Lemma~\ref{Prop:HE-type} and Proposition~\ref{Prop:EC-criteria-all} can be used in various situations.
For example, they should be useful in constructing physically reasonable models of rotating non-singular black holes or of cosmological black holes in an asymptotically expanding universe.

\subsection*{Acknowledgements}
The authors are grateful to Takuma Sato for discussions at the initial stage of the present work. 
TH is very grateful to Yu Nakayama for fruitful discussions and helpful comments.
This work was partially supported by JSPS KAKENHI Grant Numbers JP19K03876, JP19H01895 and JP20H05853 (TH).



\begin{thebibliography}{99}


\bibitem{Hawking:1971vc}
S.~W.~Hawking,
Commun. Math. Phys. \textbf{25}, 152-166 (1972)
doi:10.1007/BF01877517



\bibitem{Schon:1979rg}
R.~Schoen and S.~T.~Yau,
Commun. Math. Phys. \textbf{65}, 45-76 (1979)
doi:10.1007/BF01940959

\bibitem{Schon:1981vd}
R.~Schoen and S.~T.~Yau,
Commun. Math. Phys. \textbf{79}, 231-260 (1981)
doi:10.1007/BF01942062

\bibitem{Nester:1982tr}
J.~A.~Nester,
Phys. Lett. A \textbf{83}, 241 (1981)
doi:10.1016/0375-9601(81)90972-5

\bibitem{Witten:1981mf}
E.~Witten,
Commun. Math. Phys. \textbf{80}, 381 (1981)
doi:10.1007/BF01208277



\bibitem{Penrose:1964wq}
R.~Penrose,
Phys. Rev. Lett. \textbf{14}, 57-59 (1965)
doi:10.1103/PhysRevLett.14.57






\bibitem{Klimchitskaya:2006rw}
G.~L.~Klimchitskaya and V.~M.~Mostepanenko,
Contemp. Phys. \textbf{47}, 131-144 (2006)
doi:10.1080/00107510600693683
[arXiv:quant-ph/0609145 [quant-ph]].

\bibitem{Klimchitskaya:2009cw}
G.~L.~Klimchitskaya, U.~Mohideen and V.~M.~Mostepanenko,
Rev. Mod. Phys. \textbf{81}, 1827-1885 (2009)
doi:10.1103/RevModPhys.81.1827
[arXiv:0902.4022 [cond-mat.other]].



\bibitem{Maeda:2018hqu}
H.~Maeda and C.~Martinez,
PTEP \textbf{2020}, no.4, 043E02 (2020)
doi:10.1093/ptep/ptaa009
[arXiv:1810.02487 [gr-qc]].






\bibitem{Maeda:2021jdc}
H.~Maeda,
[arXiv:2107.04791 [gr-qc]].




\bibitem{Thakurta1983}
S. N. G. Thakurta, Indian Journal of Physics {\bf 55B}, 304 (1981).


\bibitem{Sultana:2005tp}
J.~Sultana and C.~C.~Dyer,
Gen. Rel. Grav. \textbf{37}, 1347-1370 (2005)
doi:10.1007/s10714-005-0119-7


\bibitem{McClure:2006kg}
M.~L.~McClure and C.~C.~Dyer,
Class. Quant. Grav. \textbf{23}, 1971-1987 (2006)
doi:10.1088/0264-9381/23/6/008

\bibitem{Culetu:2012ih}
H.~Culetu,
[arXiv:1202.2285 [gr-qc]].



\bibitem{Mello:2016irl}
M.~M.~C.~Mello, A.~Maciel and V.~T.~Zanchin,
Phys. Rev. D \textbf{95}, no.8, 084031 (2017)
doi:10.1103/PhysRevD.95.084031
[arXiv:1611.05077 [gr-qc]].

\bibitem{Harada:2021xze}
T.~Harada, H.~Maeda and T.~Sato,
Phys. Lett. B \textbf{833} (2022), 137332
doi:10.1016/j.physletb.2022.137332
[arXiv:2106.06651 [gr-qc]].



\bibitem{he1973}
S.W.~Hawking and G.F.R.~Ellis,
{\it ``The Large Scale Structure of Space-time''}
(Cambridge: Cambridge University Press, 1973).



\bibitem{Santos:1994cs} 
J.~Santos, M.~J.~Rebou\c{c}as and A.~F.~F.~Teixeira,
J.\ Math.\ Phys.\ {\bf 36}, 3074 (1995).
doi:10.1063/1.531013

\bibitem{srt1995} 
J.~Santos, M.~J.~Rebou\c{c}as and A.~F.~F.~Teixeira,
Gen.\ Rel.\ Grav.\ {\bf 27}, 989 (1995).
doi:10.1007/BF02113081

\bibitem{hrst1996} 
G.~S.~Hall, M.~J.~Rebou\c{c}as, J.~Santos and A.~F.~F.~Teixeira,
Gen.\ Rel.\ Grav.\ {\bf 28}, 1107 (1996).
doi:10.1007/BF02113160


\bibitem{bh2002} 
G.~Bergqvist and A.~H\"oglund,
Class.\ Quant.\ Grav.\ {\bf 19}, 3341 (2002).
doi:10.1088/0264-9381/19/12/316

\bibitem{rst2004} 
M.~J.~Rebou\c{c}as, J.~Santos and A.~F.~F.~Teixeira,
Braz.\ J.\ Phys.\ {\bf 34}, 535 (2004).
doi:10.1590/S0103-97332004000300034


\bibitem{Maeda:2020dfp}
H.~Maeda,
Gen. Rel. Grav. \textbf{53}, no.10, 90 (2021)
doi:10.1007/s10714-021-02862-8
[arXiv:2001.11335 [gr-qc]].


\bibitem{Martin-Moruno:2017exc} 
P.~Mart{\'i}n-Moruno and M.~Visser,
Fundam. Theor. Phys. \textbf{189}, 193-213 (2017)
doi:10.1007/978-3-319-55182-1\_9
[arXiv:1702.05915 [gr-qc]].



\bibitem{Lovelock:1971yv}
D.~Lovelock,
J. Math. Phys. \textbf{12}, 498-501 (1971)
doi:10.1063/1.1665613


\bibitem{Dotti:2005rc}
G.~Dotti and R.~J.~Gleiser,
Phys. Lett. B \textbf{627}, 174-179 (2005)
doi:10.1016/j.physletb.2005.08.110

\bibitem{Ray:2015ava}
S.~Ray,
Class. Quant. Grav. \textbf{32}, no.19, 195022 (2015)
doi:10.1088/0264-9381/32/19/195022


\bibitem{Ohashi:2015xaa}
S.~Ohashi and M.~Nozawa,
Phys. Rev. D \textbf{92}, 064020 (2015)
doi:10.1103/PhysRevD.92.064020



\bibitem{Gurses:1975vu}
M.~G\"urses and F.~G\"ursey,
J. Math. Phys. \textbf{16}, 2385 (1975)
doi:10.1063/1.522480






\bibitem{Burinskii:2001bq}
A.~Burinskii, E.~Elizalde, S.~R.~Hildebrandt and G.~Magli,
Phys. Rev. D \textbf{65}, 064039 (2002)
doi:10.1103/PhysRevD.65.064039
[arXiv:gr-qc/0109085 [gr-qc]].



\bibitem{Toshmatov:2014nya}
B.~Toshmatov, B.~Ahmedov, A.~Abdujabbarov and Z.~Stuchlik,
Phys. Rev. D \textbf{89}, no.10, 104017 (2014)
doi:10.1103/PhysRevD.89.104017
[arXiv:1404.6443 [gr-qc]].

\bibitem{Azreg-Ainou:2014pra}
M.~Azreg-A\"\i{}nou,
Phys. Rev. D \textbf{90}, no.6, 064041 (2014)
doi:10.1103/PhysRevD.90.064041
[arXiv:1405.2569 [gr-qc]].


\bibitem{Ghosh:2014pba}
S.~G.~Ghosh,
Eur. Phys. J. C \textbf{75}, no.11, 532 (2015)
doi:10.1140/epjc/s10052-015-3740-y
[arXiv:1408.5668 [gr-qc]].

\bibitem{Amir:2016cen}
M.~Amir and S.~G.~Ghosh,
Phys. Rev. D \textbf{94}, no.2, 024054 (2016)
doi:10.1103/PhysRevD.94.024054
[arXiv:1603.06382 [gr-qc]].



\bibitem{Torres:2016pgk}
R.~Torres and F.~Fayos,
Gen. Rel. Grav. \textbf{49}, no.1, 2 (2017)
doi:10.1007/s10714-016-2166-7
[arXiv:1611.03654 [gr-qc]].



\bibitem{Toshmatov:2017zpr}
B.~Toshmatov, Z.~Stuchl\'\i{}k and B.~Ahmedov,
Phys. Rev. D \textbf{95}, no.8, 084037 (2017)
doi:10.1103/PhysRevD.95.084037
[arXiv:1704.07300 [gr-qc]].

\bibitem{Ghosh:2020ece}
S.~G.~Ghosh, M.~Amir and S.~D.~Maharaj,
Nucl. Phys. B \textbf{957}, 115088 (2020)
doi:10.1016/j.nuclphysb.2020.115088
[arXiv:2006.07570 [gr-qc]].



\bibitem{Doran:1999gb}
C.~Doran,
Phys. Rev. D \textbf{61}, 067503 (2000)
doi:10.1103/PhysRevD.61.067503
[arXiv:gr-qc/9910099 [gr-qc]].

\bibitem{Visser:2007fj}
M.~Visser,
[arXiv:0706.0622 [gr-qc]].




\bibitem{kst1988} 
C.A.~Kolassis, N.O.~Santos, and D.~Tsoubelis,
Class. Quant. Grav. {\bf 5}, 1329 (1988).




\bibitem{Brassel:2021qxf}
B.~P.~Brassel, S.~D.~Maharaj and R.~Goswami,
PTEP \textbf{2021}, no.10, 103E01 (2021)
doi:10.1093/ptep/ptab116






\end{thebibliography}
\end{document}